\documentstyle[12pt]{article}
\begin{document}
\pagestyle{plain}
\title{\bf Missing experiments in quantum mechanics}
\author{ Miroslav Pardy\\[5mm]
Department of Physical Electronics, \\and \\
The Laboratory of the Plasma Physics,\\
Masaryk University,\\
Kotl\'{a}\v{r}sk\'{a} 2, 611 37 Brno, Czech Republic\\
e-mail:pamir@physics.muni.cz}
\date{\today}
\maketitle

\vspace{30mm}

\begin{abstract}
We discuss the two-slit experiment and the Aharonov-Bohm (AB) experiment in the magnetic field. In such a case the electron moving in the magnetic field produces so called synchrotron radiation. In other words the photons are emitted from the points of the electron trajectory and it means that the trajectory of electron is visible in the synchrotron radiation spectrum. The axiomatic system of quantum mechanics does not enable to define the trajectory of the elementary particle. The two-slit experiment and AB experiment in a magnetic field was never performed and it means that they are the missing experiments of quantum mechanics. 
The extension of the discussion to the cosmical rays moving in the magnetic field of the Saturn magnetosphere and its rings is mentioned. It is related to the probe CASSINI.
The solution of the problem in the framework of the hydrodynamical model of quantum mechanics and the nonlinear quantum mechanics is also mentioned.
 
\end{abstract}

\newpage

\section{Introduction}
\hspace{3ex}

The electron moving in the magnetic field produces so called synchrotron radiation. The production can be described by the classical electrodynamics, or, in the framework of the quantum electrodynamics (Pardy, 2007). From the view point of quantum electrodynamics, the photons are emitted from the  points of the trajectory of an electron and the trajectory of electron is visible in the synchrotron radiation spectrum. The emission of photons is stochastic and it causes that the trajectory of electrons for instance in accelerators is not stable.The axiomatic system of quantum mechanics does not enable to define the trajectory of the elementary particle, nevertheless the trajectory is  physically meaningful. The visibility of the trajectory is not the exact analogue of the trajectory of the charged particle in the ionized gas of the Wilson camera, because in this device the track is generated by the charged particle and it is visible in the daily light. Electron moving in the two-slit experiment immersed in the magnetic field radiates photons and it means that the trajectory of electron is visible in the synchrotron spectrum, or as the track in the CCD camera. Such experiment was never performed and it means that this is the missing experiment of quantum mechanics. The standard quantum mechanics considers only probability of the appearance of a particle and not its trajectory. 
The extension of our discussion to the cosmical magnetic field, or, to the Saturn magnetosphere and its rings is mentioned with regard to the probe CASSINI. 

The solution of the problem in the framework of the hydrodynamical model of quantum mechanics and the nonlinear quantum mechanics is mentioned.  In the next section we discuss the two-slit experiment in the magnetic field and then  the Aharonov-Bohm effect in the magnetic field. We follow the text of the monograph by Holstein (1992).

\section{The two-slit experiment}

The electron beam is emitted from the left to the right from the source $S$ to the screen $S_{1}$ with the upper slit $A_{1}$ and the lower slit $A_{2}$. Then it continues to the screen $S_{2}$ to form the diffraction pattern. The symbolic scheme is as follows:

$$S \rightarrow S_{1} \rightarrow S_{2}.\eqno(1)$$

The corresponding classical trajectories are $SA_{1}P $, and $SA_{2}P$, $P$ being the point of the impinging electron on the screen $S_{2}$.

Assuming the dominance of the classical trajectories (Holstein, 1992), the phase along the first trajectory is as follows:

$$\exp\left(i\int_{t_{1}}^{t_{2}}dt\frac{1}{2}m{\dot{\bf x}}^{2}\right) = \exp\left(i[\int_{S}^{A_{1}} + \int_{A_{1}}^{P}]dl\frac{1}{2}mv\right)
 \approx \exp \left(i\frac{\pi d_{0}}{\lambda} + i\frac{\pi d_{1}}{\lambda}\right),\eqno(2)$$
where

$$d_{0} = SA_{1},\quad  d_{1} = A_{1}P, \quad \lambda = \frac{2\pi}{p},\eqno(3)$$
$\lambda$ being  the deBroglie wavelength. 

The phase along the second  trajectory is as follows:

$$\exp\left(i\int_{t_{1}}^{t_{2}}dt\frac{1}{2}m{{\bf \dot x}}^{2}\right) \approx \exp \left(i\frac{\pi d_{0}}{\lambda} + i\frac{\pi d_{2}}{\lambda}\right); \quad  d_{2} = A_{2}P. \eqno(4)$$

The relative phase difference between the two paths is then given by the formula

$$\exp \left(i\frac{\pi d_{2} - \pi d_{1}}{\lambda}\right). \eqno(5)$$

The wave function at the point on the screen $S$ is then 

$$\psi \approx \psi_{0}\left(1 + \exp\left[i \pi \frac{d_{2}- d_{1}}{\lambda}\right]\right).\eqno(6)$$

The intensity is

$$|\psi|^{2} \approx |\psi_{0}|^{2}4\cos^{2}\pi \frac{d_{2} - d_{1}}{2\lambda}.\eqno(7)$$

If we denote by $L$ the distance between screens, by $\delta$ the distance between slits and by $s$ the distance of the point on the screen $S_{2}$ from the axis of symmetry of the experiment, then we can write using the the Pythagoras theorem $c^{2} = a^{2} + b^{2}; c = d_{1,2}, a = L,  b = s \pm \delta/2$ :

$$d_{2} = \sqrt{L^{2} + \left(s + \frac{1}{2}\delta\right)^{2}} \approx \sqrt{L^{2} + s^{2}} + \frac{s\delta}{2\sqrt{L^{2} + s^{2}}} = 
\sqrt{L^{2} + s^{2}} + \frac{\delta}{2}\sin\theta, \eqno(8)$$
    
$$d_{1} = \sqrt{L^{2} + \left(s - \frac{1}{2}\delta\right)^{2}} \approx \sqrt{L^{2} + s^{2}} - \frac{s\delta}{2\sqrt{L^{2} + s^{2}}} = 
\sqrt{L^{2} + s^{2}} - \frac{\delta}{2}\sin\theta .\eqno(9)$$

So, we can write the intensity distribution of electron in term of the angle $\theta$ and the slit separation in the form:

$$I(\theta) = 4I_{0}\cos^{2}\left(\pi \delta \frac{\sin\theta}{2\lambda}\right),  \eqno(10)$$
where $I_{0}$ is the intensity in case that only single slit is open.

If we switch on the magnetic field between the screen $S$ and $S_{2}$ then it is evident that the phase obtained along the first trajectory and the second trajectory is as follows:

$$(phase)_{1} \approx  \exp \left(i\frac{\pi d_{0}}{\lambda} + i\frac{\pi d_{1}}{\lambda} + ie\int_{1}{\bf A}\cdot d{\bf x} \right)\eqno(11)$$ 
and

$$(phase)_{2} \approx \exp \left(i\frac{\pi d_{0}}{\lambda} + i\frac{\pi d_{2}}{\lambda} + ie\int_{2}{\bf A}\cdot d{\bf x}
\right).\eqno(12)$$ 

The phase difference between two trajectories is evidently as follows 

$$\Delta (phase) \approx \exp \left(i\frac{\pi d_{2}}{\lambda} - i\frac{\pi d_{1}}{\lambda} + ie\oint{\bf A}\cdot d{\bf x}
\right).\eqno(13)$$

Using the Stokes theorem 

$$ie\oint{\bf A}\cdot d{\bf x} = ie\int_{area}{\rm rot}{\bf A}\cdot d{\bf S} = ie\int_{area}{\bf B}\cdot d {\bf S} = ieBS_{area},\eqno(14)$$ 
where ${\bf B}$ is the density of the magnetic induction at the area $SA_{2}PA_{1}S$.

The intensity patterns is then shifted due to the existence of the magnetic field as  

$$I(\theta) = 4I_{0}\cos^{2}\left(\pi \delta \frac{\sin\theta}{2\lambda} + e\frac{1}{2}BS_{area}\right). \eqno(15)$$

During the calculation we used the assumption that the magnetic field is sufficiently small to not change the original dominant trajectories and it is very small, to cause the change of these trajectories by the synchrotron bremsstrahlung. The synchrotron radiation is the crucial effect to see the electron trajectory in reality. Such approach can be used not only in the quantum mechanical two-slit experiments but also in case of  detections of the charged particle in the particle laboratories, or as the synchrotron radiation observation of the cosmical particles when moving in the cosmical magnetic fields. To our knowledge, such experimental approach was not applied till this time. 

\section{Aharonov-Bohm effect}

In case of the so called Bohm-Aharonov effect an infinite solenoid is introduced between slits $A_{1}, A_{2}$ on the right side of the screen ${S_{1}}$. Since the solenoid is infinite, there is no magnetic field outside the solenoid volume itself. However, There is non-vanishing vector potential outside of the solenoid, which can be expressed in the cylindrical coordinates $r, \varphi, z$ as 

$${\bf A} = \left\{\begin{array}{ll}
A_{\varphi} = \frac{1}{2}Br; \quad r < R,\\
A_{\varphi} = \frac{1}{2}B\frac{R^{2}}{r}, \quad r > R
\end{array}\right.\eqno(16)$$
which corresponds to the magnetic field nonzero inside the solenoid and zero out of the solenoid of the diameter $2R$, or,

$${\bf B} = {\rm rot}{\bf A} = B_{z} = \frac{1}{r}\frac{\partial}{\partial r}(rA_{\varphi})= \left\{\begin{array}{ll}
B_{z} = B; \quad r < R,\\
B_{z} = 0 \quad r > R
\end{array}\right.\eqno(17)$$

The phase difference at the point P between two paths is analogical to the previous discussion

$${\rm phase}\; {\rm difference} \approx \exp \left(i\frac{\pi d_{2}}{\lambda} - i\frac{\pi d_{1}}{\lambda} + ie\oint{\bf A}\cdot d{\bf x}\right)\eqno(18)$$
where the circle of integration is $SA_{2}PA_{1}S$.

According to the Stokes theorem we get in analogy with eq. (13): 

 $$ie\oint{\bf A}\cdot d{\bf x} = ie \int_{area}{\rm rot}{\bf A}\cdot d{\bf S} = ie \int_{area}{\bf B}\cdot d {\bf S} = ie BS_{area},\eqno(19)$$ 
 
 The intensity pattern is shifted according in analogy with eq. (14). Or,
 
$$I(\theta) = 4I_{0}\cos^{2}\left(\pi \delta \frac{\sin\theta}{2\lambda} + e\frac{1}{2}B\pi R^{2}\right). \eqno(20)$$

In case that in the Bohm-Aharonov arrangement the external magnetic field is switch on of the intensity $B_{ext}$, then the total shift of the intensity pattern is obviously:

$$I(\theta) = 4I_{0}\cos^{2}\left(\pi \delta \frac{\sin\theta}{2\lambda} + e\frac{1}{2}\pi B_{solenoid}R^{2} + e\frac{1}{2}B_{ext}S_{area}\right) \eqno(21)$$
where in this case $S_{area}$ is the area $SA_{2}PA_{1}S$

While the Bohm-Aharonov effect was verified experimentally, the Bohm-Aharonov effect in the homogeneous magnetic field was, to our knowledge, never performed. Also in this experiment the synchrotron radiation of electron is produced and it means that the trajectories are visible in the synchrotron radiation spectrum. Of course, we can derive the AB effect completely in terms of magnetic field and not in terms of the vector potential. However, with the same result. It is necessary only to stress that the wave function of electron "feels" the magnetic field inside the solenoid and it respects it.

\section{Discussion}

We have considered the two-slit experiment and the AB experiment in the magnetic field. In such a case the electron moving in the magnetic field produces so called synchrotron radiation. In other words, the photons are emitted from the the points of the trajectory of electron and it means that the trajectory of electron is visible in the synchrotron radiation spectrum. The axiomatic system of quantum mechanics does not enable to define the trajectory of the elementary particle, nevertheless the trajectory is physically meaningful.  Electron is an elementary particle with the localized mass and charge and it is geometrically point-like. The two-slit experiment and AB experiment in a magnetic field was never performed and it means that they are the missing experiments of quantum mechanics. The situation can be considered as the analog of the particle detection in the Wilson camera where the track of a charged particle is visible. Trajectories of the particles moving in the magnetic field  are visible in the synchrotron radiation spectrum detected by the CCD camera.

There is is a dual experiment where the magnetic monopole moves in the electric field and produces the synchrotron radiation. To our knowledge, duality was developed in particle physics but the dual experiments with magnetic monopoles were never performed. So, the dual experiments are also missing ones in quantum electrodynamics. 

The analysis can be extended to the cosmical space, where the charged cosmical rays move in the magnetic field and produce the synchrotron radiation spectrum which enables to make the trajectories visible. The opportunity was given to the cosmical probe CASSINI
moving on the orbit around the Saturn in its magnetosphere and in the magnetic field of its rings. 
The charged cosmical rays moving in such  magnetic fields generate the synchrotron radiation which can be detected by CCD camera.
In other words the trajectories are visible. So, the Auger Argentina cosmic rays project can be supplemented by the Saturn cosmical ray project realized by the probe CASSINI (Matthews et al., 2004).   
     
The problem of the visibility of trajectories can be also solved in the framework of the so called hydrodynamical model of quantum mechanics. According to Madelung (1926) Bohm and Vigier (1954),
Wilhelm (1970), Rosen (1974, 1986) and
others, the original Schr\"{o}dinger equation can be transformed into the
hydrodynamical system of equations by using the so called Madelung ansatz:

$$\Psi={\sqrt n}\*e^{\frac{i}{\hbar}\*S},\eqno(22)$$
where $n$ is interpreted as the density of particles and $S$ is the classical
action for $\hbar\rightarrow 0$. The mass density is defined by relation
$\varrho=n\*m$ where $m$ is mass of~a~particle.

It is well known that after insertion of the relation (22) into the
original Schr\"{o}dinger equation

$$i \hbar \frac {\partial \Psi}{\partial t} =
 -  \frac {\hbar^2}{2m}\*\Delta \Psi + V\* \Psi,\eqno(23)$$
where $V$ is the potential energy, we get, after separating the real and
imaginary parts, the following system of equations:

$$\frac {\partial S}{\partial t} + \frac {1}{2m}\* (\nabla\* S)^2 + V =
\frac {\hbar^2}{2m} \frac {\Delta \sqrt{n}}{\sqrt{n}}\eqno(24)$$

$$\frac {\partial n}{\partial t} + {\rm div}(n\*{\bf v}) = 0 \eqno(25)$$
with

$${\bf v}=\frac {\nabla S}{m}. \eqno(26)$$

Equation (24) is the Hamilton-Jacobi equation with the additional term

$$V_q = - \frac {\hbar^2}{2m} \frac {\Delta \sqrt{n}}{\sqrt{n}}, \eqno(27)$$
which is called the quantum Bohm potential and equation (25) is the
continuity equation.

After application of operator $\nabla$ on eq. (24), it can
be cast into the Euler hydrodynamical equation of the form:

$$\frac{\partial {\bf v}}{\partial t}+({\bf v} \cdot \nabla)\* {\bf v}=
- \frac {1}{m}\*\nabla\* (V+V_q). \eqno(28)$$

In case of the existence of magnetic field, the nonlinear equation (24), or,  (28) must be generalized for the vector potential ${\bf A}$ 
(Pardy, 2001) and then applied to the two-slit experiment and the AB experiment.

In case of the nonlinear Schr\"{o}dinger equation with the logarithmic
nonlinearity the basic equation is of the form (Pardy, 2001),
(Bialynicky-Birula et al., 1976):
  
$$i \hbar \frac {\partial \Psi}{\partial t} =
 -  \frac {\hbar^2}{2m}\*\Delta \Psi + V\* \Psi + b(\ln|\Psi|^2)\Psi,
\eqno(29)$$
where $b < 3\times10^{-15}eV$ (G\"{a}hler et al., 1981) is some constant. 

The quantum hydrodynamical equation with the nonlinear term is then 

$$m \left( \frac{\partial {\bf v}}{\partial t} + ({\bf v}\cdot\nabla)\*
{\bf v}\right)=  \nabla\left(\frac
  {\hbar^2}{2m}\frac {\Delta \sqrt{n}}
{\sqrt{n}}\right) + b(\ln|n|^2) .\eqno(30)$$
Or, 

$$ \left( \frac{\partial {\bf v}}{\partial t} + ({\bf v}\cdot\nabla)\*
{\bf v}\right)=  \nabla\left(\frac{\hbar^2}{2m^{2}} 
\frac {\Delta\sqrt{\varrho}}{\sqrt{\varrho}}\right) +
\frac{b}{m}(\ln\left|\frac{\varrho}{m}\right|^2) .\eqno(31)$$

It is evident that to find the quantum hydrodynamical  solutions
will be more complicated than of the linear quantum mechanical equation. Let us
first remember the one-dimensional solutions of the one-dimensional
nonlinear Schr\"{o}dinger equation (Pardy, 2001).

Let be $c, ({\rm Im} \;c =0), v, k, \omega$ some parameters and 
let us insert function

$$\Psi (x,t)= c\* G (x-v\*t)\* e^{i\*k\*x-i\*\omega\*t}\eqno(32)$$
into the one-dimensional equation (29) with $V=0$. Putting the imaginary
part of the new equation to zero, we get

$$v= \frac{\hbar\*k}{m}\eqno(33)$$
and for function $G$ we get the following nonlinear equation (symbol
$'$ denotes derivation with respect to $\xi= x-vt)$:

$$ G'' + A\*G + B(\ln{G})G = 0,\eqno(34)$$
where
$$A= \frac{2m}{\hbar}\*\omega - k^2 + \frac{2m}{\hbar^2}\*b \*\ln{c^2}
\eqno(35)$$

$$B= \frac{4mb}{\hbar^2}.\eqno(36)$$

After multiplication of eq. (34) by $G'$ we get:

$$\frac{1}{2}\*{\left[ G'^2 \right]}^{'} + \frac{A}{2}\*{\left[ G^2\right]}^{'}
+  B\* {\left[ \frac{G^2}{2} \ln{G} - \frac{G^2}{4} \right]}^{'} =0,\eqno(37)$$
or, after integration

$$G'^2=- AG^2 - BG^2 \ln{G} + \frac{B}{2}\* G^2 + const.\eqno(38)$$

If we choose the solution in such a way that $G(\infty)=0$ and
$G'(\infty)=0$, we get $const.=0$ and after elementary operations we get the
following differential equation to be solved:

$$\frac{dG}{G \sqrt{a-B\*\ln{G}}}= d\xi,\eqno(39)$$

where
$$a=\frac{B}{2} - A.\eqno(40)$$

Equation (39) can be solved by the elementary integration and the
result is

$$G= e^{\frac{a}{B}}\*e^{-\frac{B}{4}\*(\xi+d)^2},\eqno(41)$$
where $d$ is some constant.

The corresponding soliton-wave function is evidently in the one-dimensional
free particle case of the form:

$$\Psi(x,t)= 
c\*e^{\frac{a}{B}}\*e^{-\frac{B}{4}\*(x-vt+d)^2}\*e^{ikx-i\omega\*t}.
\eqno(42)$$

It is not necessary to change the standard probability interpretation of the
wave function. It means that the normalization condition in our one-dimensional
case is

$$\int_{-\infty}^{\infty}{\Psi^*\*\Psi\,dx} =1.\eqno(43)$$

Using the Gauss integral

$$\int_{0}^{\infty}{e^{-\lambda^2\*x^2}\,dx} =
\frac{\sqrt{\pi}}{2\lambda},\eqno(44)$$
we get with $\lambda= {\left(\frac{B}{2}\right)}^{\frac{1}{2}}$

$$c^2\*e^{\frac{2a}{B}}= {\left(\frac{B}{2\pi}\right)}^{\frac{1}{2}}\eqno(45)$$
and the density probability $\Psi^*\Psi = \delta_m(\xi) $ is of the form
(with $d=0$):

$$\delta_m(\xi)= \sqrt{\frac{m\alpha}{\pi}}\* e^{-\alpha m \xi^2}
\hspace{5mm};\hspace{5mm}\alpha= \frac{2b}{\hbar^2}.\eqno(46)$$

It may be easy to see that $\delta_m(\xi)$ is the delta-generating function and
for $m \rightarrow \infty$ is just the Dirac $\delta$-function.

It means that the motion of a particle with sufficiently big mass $m$ is
strongly localized and in other words it means that the motion of this particle
is the classical one (There are no quantum jumps of the Moon). Such behavior of a particle cannot be obtained in the
standard quantum mechanics because the plane wave
$\exp[ikx-i\omega\*t]$
corresponds to the free particle with no possibility of localization
for $m \rightarrow \infty$.

Let us still remark that it is possible to show that coefficient $c^2$ is real and positive number (Pardy. 2001). The generalization of the equation (29), or, (30) can be performed by involving the vector potential ${\bf A}$ into them for the solution of the two-slit experiment and AB experiment in the presence of the magnetic field (Pardy, 2001).
 
We frequently read in the physical texts on the quantum mechanics that 
the classical limit of quantum mechanics is obtained only by the so called
WKB method. However, the limit is only formal because in this case
the probabilistic form of the solution is conserved while classical
mechanics is strongly
deterministic. In other words, statistical description of quantum 
mechanics is in no case reduced to the strong determinism of 
classical mechanics of one-particle system. So, only nonlinear 
quantum mechanics of the above form gives the correct classical 
limit expressed by the delta-function. More information on the
problems which are solved by the nonlinear  Schr\"{o}dinger  equation
involving the collapse of the wave function and the Schr\"{o}dinger
cat paradox is described in author's articles (Pardy, 2001, 1994). The
extended version of the nonlinear  quantum world is described
in the preprint of Castro (2002).

\newpage

\noindent
{\bf References}

\vspace{5mm}

\noindent
Bialynicky-Birula, I. and  Mycielski, J., Nonlinear wave mechanics, Ann. Phys. (N.Y.) 100 (1976) 62.\\[2mm]
Bohm, D. and  Vigier, J.,  Model of the causal interpretation of
quantum theory in terms of a fluid with irregular fluctuations, Phys. Rev. 96 No. 1 (1954) 208.\\[2mm]
Castro, C., Mahecha, J. and Rodr\'iguez, B.,
Nonlinear QM as a fractal Brownian motion with complex diffusion constant, quant-ph/0202026.\\[2mm]
G\"{a}hler, R.,  Klein,  A. G. and  Zeilinger, A., Neutron optical
test of nonlinear wave mechanics,  Phys. Rev. A 23 No. 4 (1981) 1611.\\[2mm]
Holstein, B. R., Topics in advanced quantum mechanics, Addison-Wesley Publishing Company, Redwood City, (1992). \\[2mm]
Madelung, E., Quantentheorie in hydrodynamischer Form, Z. Physik 40 (1926) 322.\\[2mm]
Matthews, L. S. and Hyde, T. W., Gravitoelectrodynamics in Saturn's F Ring: Encounters with Prometheus and Pandora,
preprint of the Center for Astrophysics, Space Physics, and Engineering Research, Baylor University, (2004). \\[2mm]
Pardy, M., Possible Tests of Nonlinear Quantum Mechanics,
in: Waves and Particles in Light and Matter, Ed. by Alwyn van der Merwe and Garuccio, A., Plenum Press New York (1994).\\[2mm]
Pardy, M., To the nonlinear quantum mechanics, quant-ph/0111105.\\[2mm]
Pardy, M., The synchrotron radiation from the Volkov solution of the Dirac equation: e-print hep-ph/0703102. \\[2mm]
Rosen, N., A classical picture of quantum mechanics, Nuovo Cimento 19 B  No. 1 (1974) 90.\\[2mm]
Rosen, N., Quantum particles and classical particles, Foundation of Physics 16 No. 8 (1986) 687.\\[2mm]
Wilhelm, E., Hydrodynamic model of quantum mechanics, Phys. Rev. D 1 No. 8 (1970) 2278.

\end{document}